\documentclass[iop]{emulateapj}

\usepackage{amsmath}
\usepackage{subfigure}
\usepackage{xfrac}
\usepackage{amsfonts}
\usepackage{graphicx}
\usepackage[dvipsnames]{xcolor}
\usepackage{mathtools}
\usepackage{afterpage}
\usepackage{nicefrac}
\usepackage{listings}
\usepackage{wasysym}
\usepackage{ifsym} 
\usepackage{url}
\makeatletter
\newcommand\Label[1]{&\refstepcounter{equation}(\theequation)\ltx@label{#1}&}
\makeatother

\usepackage{times}

\newcommand{\beq}{\begin{equation}}
\newcommand{\eeq}{\end{equation}}
\newcommand{\bea}{\begin{eqnarray}}
\newcommand{\eea}{\end{eqnarray}}

\newcommand{\wave}{_w}
\newcommand{\m}{_{\rm max}}

\newcommand{\shock}{_{\rm s}}

\newcommand{\pt}{\partial}

\newcommand{\msun}{\mathrm{M}_\odot}

\newcommand{\Lwmax}{L_{{\rm max}}}

\newcommand{\calR}{\mathcal{R}}

\newcommand{\tti}{\tilde{t}_i}
\newcommand{\sinc}{{\rm sinc}}

\newcommand{\inv}{^{-1}} 
\newcommand{\kmpsec}{km\,s$\inv$}

\newcommand{\sro}{\color{Sepia}}

\newcommand{\software}{{\em Software:}} 
\newcommand{\cdmrev}{} 
\shortauthors{Matzner \& Ro}
\shorttitle{Wave-Driven Shocks}

\begin{document}

\title{Wave-Driven Shocks in Stellar Outbursts: Dynamics, Envelope Heating, and Nascent Blastwaves}

\author{Christopher D. Matzner$^1$ \& Stephen Ro$^2$\email{sro@berkeley.edu}}
\affil{$^1$Department of Astronomy \& Astrophysics, University of Toronto, 50 St. George St., Toronto, ON M5S 3H4, Canada}
\affil{$^2$Astronomy Department and Theoretical Astrophysics Center, University of California, Berkeley, Berkeley, CA 94720}
\received{September 25, 2020} 
\begin{abstract}
{ We address the shocks from acoustic pulses and wave trains in general one-dimensional flows, with an emphasis on the application to super-Eddington outbursts in massive stars.   Using approximate adiabatic invariants, we generalize the classical equal-area technique in its integral and differential forms.  We predict shock evolution for the case of an initially sinusoidal but finite wave train, with separate solutions for internal shocks and head or tail shocks, and demonstrate detailed agreement with numerical simulations.  Our internal shock solution motivates improved expressions for the shock heating rate.  Our solution for head and tail shocks demonstrates that these preserve  dramatically more wave energy to large radii and have a greater potential for the direct ejection of matter.  This difference highlights the importance of the waveform for shock dynamics.  Our weak-shock analysis predicts when shocks will become strong and provides a basis from which this transition can be addressed. {\cdmrev We use it to estimate the mass ejected by sudden sound pulses and weak central explosions. }  }
\end{abstract}

\section{Introduction}
\label{sec:introduction}
A fraction of massive stars undergo periods of intense, episodic mass loss as they approach core collapse.  Observations of eruptions in luminous blue variables, supernova (SN) impostors, and  pre-SN outbursts provide direct evidence of these events \citep{2007ApJ...657L.105F, 2007Natur.447..829P, 2014ApJ...780...21M, 2014MNRAS.438.1191S}, while dense and massive zones of circumstellar media (CSM) provide indirect evidence.  Dense CSM is revealed by its interaction with supernova (SN) ejecta \citep{2008MNRAS.389..113P, 2011MNRAS.412.1522S, 2012ApJ...744...10K,2014MNRAS.439.2917M, 2016MNRAS.456..853P, 2016MNRAS.461.3057S, 2017ApJ...836..158H,2017MNRAS.471.4381S}, or  by recombination after `flash ionization' from SN shock breakout \citep{2014Natur.509..471G, 2014A&A...572L..11G, 2016ApJ...818....3K}.  SN types IIn and Ibn require especially dense CSM from previous eruptions \citep{Chugai04,Pastorello08a-Herich}, and SN spectral evolution from Type Ib to IIn \citep{2015ApJ...815..120M,2017ApJ...835..140M} and the transitional Type IIn/Ibn SNe \citep{2008MNRAS.389..131P,2012MNRAS.426.1905S} show strong interactions with dense, hydrogen-rich CSM.

Importantly,  ejecta speeds in some outbursts are high enough to require acceleration by shocks, as opposed to winds.   
The pre-explosion outburst of SN 2009ip emitted material at 13,000\,\kmpsec \citep{Pastorello13_09ip}, and there is tentative evidence of 15,000\,\kmpsec ejecta in the similar source SN 2015bh \citep{Ofek16_2015bh}.
The famous Galactic example $\eta$\,Car has survived multiple eruptions; its last Giant Eruption ejected traces material at SN velocities (${\sim}20\,000$\,\kmpsec; \citealt{2018MNRAS.480.1457S}) along with about ${\sim}15$\,$\msun$ of slower (${\sim}600$\,\kmpsec) material over two decades \citep{2008Natur.455..201S}.


{  Such phenomena require that energy escapes the stellar interior by a conduit capable of transporting super-Eddington luminosities and creating shock motions along the way.  Sound waves are a prime suspect for this conduit, for which the energy sources may include unstable phases of oxygen-silicon shell burning in low-mass pre-SN stars \citep{2015ApJ...810...34W} or pulsational pair instability in high-mass pre-SN stars \citep{2007Natur.450..390W}}. 
More commonly, waves from vigorous convection in advanced burning regions \citep{2012MNRAS.423L..92Q, 2014ApJ...780...96S, 2014ApJ...785...82S} can  transport energy to the outer envelope at super-Eddington rates.  Wave dissipation is then thought to trigger outflows with mass loss rates and velocities characteristic of Type IIn SNe CSM \citep{2016MNRAS.458.1214Q,2017MNRAS.470.1642F,2018MNRAS.476.1853F}.   

{Thanks to the pulsational pair instability, wave-driven mass loss plays a role in determining the lower edge of an upper black hole mass gap \citep{woosley2017pulsational,LeungNomotoBlinnikov19,Farmer19_PPI}, a phenomenon that is currently of interest due to the high inferred initial masses in the black hole coalescence GW190521 \citep{LIGO2020_GW190521}.

We seek a way to accurately predict the formation and evolution of wave-driven shocks and the distribution of shock-deposited heat.  Critically, a complete theory must also evaluate the conditions under which some of the wave energy persists all the way to the stellar surface, driving an eruption.  Closed-form solutions are especially useful: while waves and shocks can be calculated using hydrodynamical codes, the very wide range of physical scales makes direct simulation very costly in practice.  Thousands of cells are needed per acoustical wavelength even with high-order codes, and the wavelength may be very small compared to the stellar radius.     

Fortunately, very accurate predictions are possible using an analytical technique, which in fact is most accurate in the same  (low-amplitude and high-frequency) limit in which direct numerical simulations are the least feasible.  It is also a classic technique, as it builds on a proposal by \citet{lighthill78} for the generalization of a method due to \citet{whitham1974linear}, which in turn follows a suggestion of \citet{landau1945shock} on how to extend the seminal theory by \citet{Riemann1860} past the point of shock formation.  We start here by tracing these contributions, and reviewing our own work on the validity of \citeauthor{lighthill78}'s proposal.  We then extend the theory somewhat by re-expressing it in differential form, before considering its implications for shock dissipation within stars.  Our results  improve upon, and generalize, the shock heating prescription developed by \citet{2018MNRAS.476.1853F}.   We shall see that the potential for shock emergence depends critically on waveform of the driving source, as well as its amplitude and period.

{\em Notation}: Our space and time coordinates are $r$ and $t$; $r$ will be the stellar radius when we consider spherical symmetry.  We use $\rho$, $c$, $p$, $\gamma$, $v$, and $A$ for the density,  sound speed, pressure, adiabatic index, fluid velocity, and cross-sectional area.  For simplicity we consider only ideal gases, in which $\gamma$ is fixed in each fluid element (but may vary across the structure); and we define the auxiliary quantity $q=(\gamma-1)/2$.  The prefix $\delta$ indicates perturbations from the background state, which is indicated by subscript $0$.  The jump of a quantity across a shock is denoted by prefix $\Delta$. 

\section{The equal-area method and its generalization}

The simplest problem in nonlinear fluid dynamics involves a constant-entropy fluid moving within a channel of constant cross-sectional area (or in planar symmetry), in the absence of viscosity or any body forces.  This problem was solved exactly by \citet{Riemann1860}, who  re-expressed the fluid equations in terms of quantities $J_\pm = v \pm 2c/(\gamma-1)$ that are conserved along sound fronts: outward or inward trajectories defined by $\dot{r} = v\pm c$, also known as  $C^\pm$ characteristics.    \citeauthor{Riemann1860}'s solutions apply only while the flow is continuous; however, they generically produce discontinuities after some finite time.  This is most obvious for `simple' waves, in which one of the two wave families is constant: for instance, if $J_- = 2 c_0/(\gamma-1)$ for some constant $c_0$, then $v = 2(c-c_0)/(\gamma-1) = J_+/2$ and the wave speed $v+c = (\gamma+1)v/2 + c_0$ is a pure function of $J_+$ or of $v$.  This implies that the velocity waveform $v(r-c_0 t)$ will shear at a strictly uniform rate, as shown in Figure \ref{fig:EqualArea}, with wavefronts moving horizontally (at constant $v$).   Within this shearing waveform, the  inverse velocity gradient $(\partial v/\partial x)^{-1}$ increases linearly with time, at the uniform rate $(\gamma+1)/2$, along each wavefront.  Therefore, compressive regions of the waveform -- those in which $\partial v/\partial r$ is initially negative -- will develop infinite slope in a finite time.  This happens first on whichever sound front or fronts carry the greatest compression, at a time $t_s = [2/(\gamma-1)] \min (-\partial v/\partial x)_{t=0}^{-1}$.  After $t=t_s$, \citeauthor{Riemann1860}'s solution requires modification because it predicts a shearing waveform that has `broken' and become multi-valued, whereas in reality the flow is single-valued and contains shock discontinuities.  

The `equal-area' method \citep[][following \citealt{landau1945shock}; see also \citealt{landau1959fluid} \S\,102]{whitham1974linear} provides the modified waveform, including the location and strength of the shocks that form.  One assumes that \citeauthor{Riemann1860}'s solution remains valid away from the shock discontinuities;  shocks simply remove a portion of the waveform.   Mass and momentum conservation require that the removed sections of $v(r)$ must be of equal area, and this determines the shock's location as a function of time.   Because  the waveform shears in a simple and uniform way, as do the omitted areas, one can identify the shock transition at time $t$ directly from the initial waveform (at $t=0$), simply by determining a line of appropriate slope that cuts away sections of equal area.  Figure \ref{fig:EqualArea} provides an example. 

Once the strength of each shock is known, there are two ways to obtain the rate at which it transfers wave energy ($E_w$) into heat ($E_{\rm diss}$) within the stellar envelope.  One is to employ the local formula  
\begin{equation}\label{eq:dEwdr-dissipation}
\frac{dE_{\rm diss}}{dr} = - \frac{dE_w}{dr} 
=   \frac{\gamma+1}{12} A \rho_0 \frac{(\Delta v)^3}{c_0}, 
\end{equation} 
where $\Delta v$ is the velocity jump across the shock.  The other involves inspecting the waveform to determine how much of its total energy has been removed by the excision of sections around the shock discontinuities. 

While \citeauthor{Riemann1860}'s solution is exact and nonlinear, this cannot be said for the equal-area method.  However, its approximations are highly accurate so long as the wave amplitude is small ($v\ll c_0$) and the shocks are correspondingly weak.   A sufficiently weak shock adds negligible entropy to the fluid, as the entropy perturbation is of order the shock strength cubed; nor does it create a significant reflected wave (change in $J_-$), either of  which would ruin \citeauthor{Riemann1860}'s solution for the fluid downstream of the shock.   In the small-amplitude limit, moreover, the perturbations of the conserved mass density $\rho$ and momentum density $\rho v$ are proportional to that of $v$, so the equal-area constraint applies to any of these waveforms.

 \begin{figure}[ht]
\includegraphics[width=0.5\textwidth]{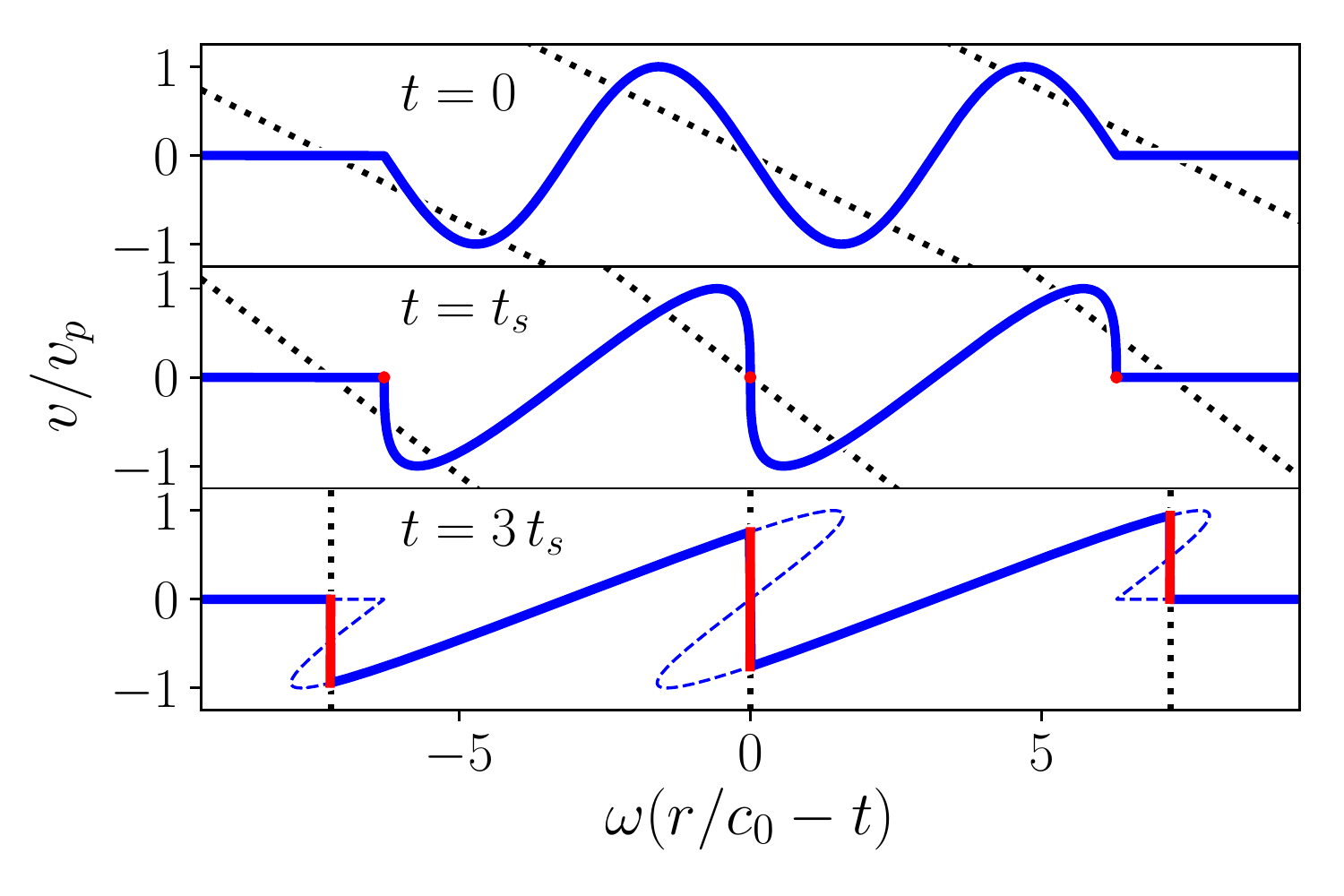}
\caption{ 
\citeauthor{whitham1974linear}'s equal-area technique applied to a two-cycle outward-traveling sinusoidal wavetrain of amplitude $v_p$, in the simple problem of an isentropic fluid within a constant-area channel.  The invariance of $J_+$ along $C^+$ characteristics leads the waveform to shear at a uniform rate, so shocks form within the waveform at the locations of maximal $-\partial v/\partial x$ at time $t=t_s$ (red dots).   Shock transitions at time $t$ shear back to lines in the initial conditions (black dotted lines, corresponding to shocks at $t=3t_s$) that cut away equal positive and negative areas of the initial waveform (dashed blue regions).  Internal shocks evolve more rapidly than head and tail shocks, with implications for energy deposition and shock ejection. 
}
\label{fig:EqualArea}
\end{figure} 

\subsection{\citeauthor{lighthill78}'s generalization} \label{SS:Lighthill_Method}

In general, unlike the simple problem, the cross-sectional area, fluid density, and sound speed may all vary with $r$.  A further modification is thus required to adapt the equal-area method to more complicated environments like stellar envelopes, with their spherical geometries and strongly stratified density profiles. For this,  \citeauthor{lighthill78} ({\em Waves in Fluids}, \citeyear{lighthill78}, chapter 2) proposed to generalize \citeauthor{whitham1974linear}'s method by replacing $J_\pm$ with an approximate invariant (a linearized adiabatic invariant, to be precise) of acoustical motions in the more complicated problem.   \citeauthor{lighthill78} chose the approximate invariant $\delta p/\sqrt{Z}$, where $\delta p$ is the pressure perturbation and $Z(r) = \rho_0 c_0/A$ is the acoustic impedance.  However we prefer to use the `wavefront amplitude', 
\begin{equation} 
\calR_\pm = \sqrt{\rho_0 c_0 A} ~~\delta J_\pm
\end{equation} 
\citep[][hereafter MR20]{2019jfm}.  Here, $\delta J_\pm$ is the acoustical perturbation of $J_\pm$ from the background state.  Our quantity $\calR_\pm$ is also an approximate invariant.  Its advantage is that it provides a more direct connection to $J_\pm$ and also to the acoustical wave luminosity, which for waves of linear amplitude is 
\begin{equation}
L_w = \frac14 (\calR_+^2 - \calR_-^2).
\end{equation}
  Conservation of $\calR_+$ along characteristics in a purely outward-traveling wave ($\calR_-=0$) is therefore equivalent to the conservation of $L_w$ in a wave that neither reflects nor dissipates. 

In the generalized method, the wavefront shears at a changing rate that depends on the local conditions.   This can be accommodated by defining a new evolution coordinate $y(r)$ to replace $t$, as we describe in more detail in \S~\ref{S:differential_form}.  Before we do, we wish to review a couple results from MR20 that help to clarify the validity of the generalized method. 

First, it is important to note that the quantity $\calR_\pm$ is not strictly conserved along characteristics, but rather varies in linear theory according to 
\beq \label{eq:calReqn} 
\frac{d}{dr^\pm} \calR_\pm =\calR_\mp \frac{\partial}{\partial r} \ln \sqrt{Z}
\eeq
where $d/d r^\pm$ is the derivative along $C^\pm$ with respect to $r$.   The right-hand side causes reflection whenever the impedance $Z$ varies significantly on scales of order the wavelength, but is completely negligible when it does not (the high-frequency limit).  Therefore, \citeauthor{lighthill78}'s generalized equal-area method requires that the wave  not only be linear in amplitude, but high in frequency relative to its surroundings.   This is not a strong constraint in the {\cdmrev process of shock formation}, because the dynamical timescales are much shorter in the core than in the envelope of an evolved star, and because a wave must be low-amplitude to travel many wavelengths before it forms a shock.   However, if the wave were to meet a rapid change in $Z$ (at a composition boundary, for instance), then one would need to evaluate what fraction is transmitted and apply our technique to the transmitted wave.  {\cdmrev Reflection may also affect shocks that accelerate near the stellar surface, as we discuss in \S~\ref{SS:mass-ejection}. } 

Second, equation (\ref{eq:calReqn}) shows that $\calR_\pm$ is an adiabatic invariant of the linearized problem, but shock formation is a nonlinear process.   In MR20 we show that nonlinear adiabatic invariants, when these can be found, almost always tend toward $\calR_\pm$ in the linear limit. More generally, we find that the nonlinear error one incurs by assuming  that $\calR_\pm$ is constant is second-order in the wave amplitude, and also proportional to the logarithmic gradient of a combination of environmental quantities.  Note also that the `conserved area' $\int \calR_+(t_i) dt_i$ has units of the square root of the action, which should be conserved in adiabatic evolution.   These results give us confidence that  \citeauthor{lighthill78}'s proposed generalization is sound. 


\section{The  method in differential form} \label{S:differential_form} 

Here we shall present an analysis that is physically equivalent to the equal-area method, but expressed in differential rather than integral form.  As before, we focus on outward-traveling waves that encounter no inward sound energy, and that also are sufficiently high-frequency to suffer negligible reflection; accordingly, $\calR_- = \delta J_-=0$.  This implies $\delta J_+=2v$ and  $L_w=A\rho_0c_0 v^2$, so  \[ \frac{v^2}{c_0^2} = \frac{L_w}{\Lwmax(r)}\] where $\Lwmax(r) = A \rho_0 c_0^3$.  

We label the $i$-th characteristic by the time  $t_i$ at which it crosses a fiducial radius $r_1$, and by its conserved amplitude $\calR(t_i)$ and  luminosity, $L_{w}(t_i) = \calR(t_i)^2/4$.  It is convenient to define the normalized waveform $f(t_i) = \calR(t_i)/\calR_{p}$     where $\calR_p$ is the amplitude at the wave peak (and correspondingly, $L_p = \calR_p^2/4$). 
Travelling at speed $(v+c)_i= c_0+(q+1)v_i$, characteristic $i$ arrives at radius $r$ at the time
\bea
t &=& t_i + \int_{r_1 }^r\frac{dr'}{(v+c)_i} \nonumber \\
&=& t_i + \int_{r_1}^r\frac{dr'}{c_0(r')} - f(t_i) y(r),
\label{eq:arrivaltime}
\eea
Above, we take the Taylor expansion of the right hand side in terms of $v_i/c_0 \ll1$ to generate the second equation, and  we define two new quantities:
\beq
y(r) = L_p^{1/2}Y(r)
\eeq
and
\beq
Y(r) = \int_{r_1}^r\frac{dr'}{c_0(r')}\frac{\gamma(r')+1}{2L\m^{1/2}(r')}.  \label{eq:Y}
\eeq
%
The evolution coordinate $y(r)$ takes the role played by $t$ in the simple problem, in that it determines how much the waveform has sheared.   
 
Since $t$ is a function of $r$ and $t_i$, the partial derivative with respect to $t_i$ commutes with the integration over $r$. The density of characteristics { is indicated by $\partial t_i/\partial t$, which is} the inverse of
\beq {\partial t}{/\partial t_i} = 1 - f'(t_i) y(r). \label{eq:pdtime1} \eeq

{ The wave variation evolves according to}
\beq
\frac{\partial f}{\partial t} = \frac{df}{dt_i}\frac{\partial t_i}{\partial t} =  \frac{f'(t_i)}{1 - f'(t_i)y(r)}. 
\label{eq:wave_gradient}
\eeq
Characteristics converge (disperse) with time { in compressive (rarefactive) regions} of a wave, since $y(r)$ is positive and increasing.  Note that $\partial f/\partial t$ and $\partial v/\partial r$ are opposite in sign: compressive regions have $\partial f/\partial t>0$. 

\subsection{Shock formation}

A shock forms where $\partial f/\partial t$ becomes infinite; this occurs first on those characteristics, labeled $t_{i,s}$, that are local maxima of the compression rate and therefore local maxima of $f'(t_i)$.  (This is a necessary but not always sufficient criterion: it is possible for such characteristics to collide with another shock before forming their own.) 
The shock formation radius $R_s$ is found by the relation 
\beq
y_s \equiv y(R\shock) = \frac{1}{f'(t_{i,s})}. \label{eq:shockformationradius}
\eeq
Nearby characteristics will collide with the shock: these correspond to the sections of the waveform excised in the equal-area method. 

The shock formation criterion, equation (\ref{eq:shockformationradius}), was derived by \citeauthor{lighthill78} (\citeyear{lighthill78}, his equation 254), and re-derived by \citet{2001JFM...431..161L}, \citet{tyagi2003nonlinear}, 
\citeauthor{tyagi2005propagation} (\citeyear{tyagi2005propagation}, who noted agreement with \citeauthor{lighthill78}'s criterion) and us \citep{2017ApJ...841....9R}.   As we discuss in MR20, it assumes the wave's slope is not altered by reflection at some location where $Z$ changes rapidly; this is more of a restriction for internal shocks than head shocks.

\subsection{Weak Shock Propagation}
\label{sec:weakshock}
To address the trajectory  $R(t)$  of a shock after it forms, we follow the characteristic-tracking procedure of \cite{friedrichs48} and \cite{whitham1974linear}. 
The velocity $V=dR/dt$ of a weak shock is just the average of the characteristic velocities immediately to each side of it:
\beq
V=\frac{(v+c)_u + (v+c)_d}{2},
\label{eq:vshockcharacteristics}
\eeq
to first order in $v/c_0$ \citep{1948sfsw.book.....C}, where $u$ and $d$ refer to the states immediately upstream and downstream of the shock, respectively.  This condition enforces the equal-area rule we discussed in \S~\ref{SS:Lighthill_Method}, but in differential form. 

{ In order to evaluate $V$, we require the rate at which characteristics arrive at the shock front from either side:}
\beq
\left. \frac{dt_i}{dt} \right|_s = \partial_t t_i + V\partial_r t_i =\frac{v+c-V}{v+c}\pt_t t_i \nonumber;
\eeq
{the second equality uses $\partial_t t_i = -(v+c)\partial_r t_i$}. 
Substituting equation (\ref{eq:pdtime1}),
\beq
\left(1 - f'(t_i)y \right)\left. \frac{dt_i}{dt} \right|_s\simeq \frac{v+c-V}{v+c}, \nonumber
\eeq
{which can be evaluated on either side of the shock.}
Combining this with equation 
(\ref{eq:vshockcharacteristics}) and using $(q+1)\sqrt{L_p/\Lwmax} \left. d/dt\right|_s = \left. d/dy\right|_s$ gives coupled differential equations for the  initial times of characteristics as they arrive at the shock:
\begin{subequations}\label{eq:shocktrajectory2}
\bea
2\left(1 - f'(t_{id}) y \right)  \frac{dt_{id}}{dy}  &\simeq& f(t_{id})-f(t_{iu}), \label{eq:shocktraj2-downstream}
\\ 
2\left(1 - f'(t_{iu}) y \right)  \frac{dt_{iu}}{dy}  &\simeq& f(t_{iu})-f(t_{id}).
\label{eq:shocktraj2-upstream}
\eea
\end{subequations}
 { The shock trajectory, described by $t_{iu}(y)$ and $t_{id}(y)$, is found by integrating these equations from the point of shock formation, $t_{iu} = t_{id} = t_{i,s}$.   
When it is desirable to work with dimensionless solutions scaled to the formation of a given shock, we note that equations  (\ref{eq:shocktraj2-downstream}) and  (\ref{eq:shocktraj2-upstream}) remain valid and unchanged with the replacements $t_i\rightarrow \tti$ and $y\rightarrow \tilde{y}$,  where 
 \[\tti \equiv \frac{t_i-t_{i,s}}{y_s}, ~~~~~ ~\tilde{y}(r) \equiv \frac{y(r)}{y_s} = \frac{Y(r)}{Y_s},\] 
and $Y_s$ means $Y(R_s)$.    

}

{ 

To specify the deposition of wave energy we define a normalized energy variable $\tilde{E} = E/(L_p y_s)$.  Equation (\ref{eq:dEwdr-dissipation}) becomes 
\begin{equation} \label{eq:dEwdr-dimensionless}
\frac{d\tilde{E}_{\rm diss}}{dr} = \frac{(\Delta f)^3}6 \frac{d\tilde{y}}{dr} = \frac{d\tilde{E}_{\rm diss}}{d\tilde{y}}  \frac{d\tilde{y}}{dr}
\end{equation} 
We see from this that shock  dissipation can be broken into two parts.  The first,  $d\tilde{E}_{\rm diss}/d\tilde{y} = (\Delta f)^3/6$, depends only on the form of the initial pulse and the evolution coordinate $\tilde{y}$.  The second, $d\tilde{y}/dr$, depends only on the wave amplitude and the structure of the background envelope.  }

\section{Particular Solutions}
\label{sec:particular_solutions}
{ 
Equations (\ref{eq:shocktraj2-upstream}) and (\ref{eq:shocktraj2-downstream}) are both of the form \[ \left( f(t_{id,iu})-f(t_{iu,id})\right) y' = 2\left(1- f'(t_{id,iu})y\right) ,\] which  is solvable if there is a definite relationship between the values of $f$ just upstream and downstream of the shock.  Suppose $f(t_{id,iu})-f(t_{iu,id})=mf(t_{id,iu})$}
for some non-zero constant $m$; then 
\beq
b\left(1 - f'(t_i) y(t_i) \right) = f(t_i)y'(t_i),
\label{eq:shocktrajectory3}
\eeq
where $b=2/m$ 
{ 
and $t_i$ is shorthand for either $t_{id}$ or $t_{iu}$.} The exact solution is
\beq
y(t_i) =  y_s 
\frac{f(t_{i,s})^b}{f(t_i)^b} + \frac{b}{f(t_i)^{b}} \int_{t_{i,s}}^{t_i} f(t_i')^{b-1}dt_i'
\label{eq:shocktrajectory_soln}
\eeq
{ The dimensionless form of this solution is exactly the same, with  $\{t_{i,s},t_i,y\}\rightarrow \{0,\tti,\tilde{y}\}$. }

{ For a shock that forms at a wave node ($f(t_{i,s})=0$), the first term in equation (\ref{eq:shocktrajectory_soln}) is zero and can be omitted. }

This solution addresses some common cases.  {\em Internal} shocks are those for which there is wave power both upstream and downstream of the shock.  When $f(t_i)$ is locally symmetric around $t_{i,s}$, we know that $t_{iu}-t_{i,s} = t_{i,s}-t_{id}$ and $f(t_{iu})=-f(t_{id})$; these shocks are described by $m=2$ and $b=1$.  Internal shocks travel at the undisturbed sound speed ($V=c_0$), to first order in the wave amplitude.   {\em Head} and {\em tail} shocks are those in which a wave packet suddenly begins or ends in a compressive phase.  For these, either $f_{id}$ or $f_{iu}$ is zero, so they are described by $m=1$ and $b=2$.   Leading shocks travel somewhat faster than $c_0$; whereas trailing shocks travel slightly slower than $c_0$ (and yet are supersonic relative to the upstream fluid).

\subsection{Shocks in Sinusoidal Waves}
\label{sec:sinusoidal}

Consider a sinusoidal wave of $N$ cycles,  $f = \sin \omega t_i$ for $0<\omega t_i < 2N\pi$.  We assume that it begins and ends in a state of compression, so that head and tail shocks form in addition to its internal shocks.   All these shocks share the common value 
$y_s = 1/\omega$, { so $\tilde{y} = \omega y$ and one can define $\tti = \omega (t_i-t_{i,s})$ for any shock labeled by $t_{i,s}$.   The example in Figure \ref{fig:EqualArea} consists of two cycles.

\subsubsection{Internal shocks} \label{SSS:InternalShocks}

Internal shocks evolve according to equation (\ref{eq:shocktrajectory_soln}) with $b=1$ and $f(t_{i,s})=0$, so 
{ $\tilde{y}(\tti) = \tti/\sin(\tti)$ and 
\beq  \tti = \sinc^{-1} \frac{1}{\tilde{y}(r)}  \label{eq:SinInt-y(ti)}.\eeq
This coincides with what one would obtain by construction for the internal shock in Figure \ref{fig:EqualArea}.

Symmetry around the internal shocks implies $u_u=-u_d$ and so $\Delta f = 2 f_d$.  Evaluating equation (\ref{eq:dEwdr-dimensionless}), 
\begin{equation} \label{eq:dEdy-sin-internal} 
\frac{d\tilde{E}_{\rm diss}}{d\tilde{y}} = \frac43 \left[ \sin\left(\sinc^{-1}\frac1{\tilde{y}}\right)\right]^3
\end{equation}
and the cumulative dissipated energy is 
\begin{equation} \label{eq:Ediss_cumulative-sin-internal} 
\tilde{E}_{\rm diss} = \tti - \sin \tti \cos\tti - \frac23 \tti (\sin \tti)^2, 
\end{equation} 
which can be converted to $\tilde{E}_{\rm diss}(\tilde{y})$ using (\ref{eq:SinInt-y(ti)}).   We obtained this by integrating the wave energy in the equal-area portions of the initial waveform that will be removed by the shock (see Figure~\ref{fig:EqualArea}); we then checked for numerical consistency with equation (\ref{eq:dEdy-sin-internal}). 

The total energy deposited by an internal shock is $\tilde{E}_{\rm diss}(\infty) = \int_1^\infty (d\tilde{E}_{\rm diss}/d\tilde{y}) d\tilde{y} = \pi$, or in  dimensional terms, $E_{\rm diss}(\infty) = \pi L_p/\omega$.  This matches the initial wave energy carried by  characteristics that will intersect the shock.  We will also be interested in what fraction of the wave energy survives at late times.  At large values of $\tilde{y}$, $d\tilde{E}_{\rm diss}/d\tilde{y} \rightarrow (4\pi^3/3)\tilde{y}^{-3}$, so the residual wave energy that will ultimately interact with the shock is 
\beq \label{eq:Ew-residual-int}
\tilde{E}_w(\tilde{y})\rightarrow \frac23 \pi^3 \tilde{y}^{-2}.
\eeq

\subsubsection{Head and tail shocks} \label{SSS:HeadTailShocks}

Head and tail shocks evolve in opposite directions but otherwise in exactly the same way (in linear theory), so we focus on the case of a head shock.  Using $b=2$ and $f(t_{i,s})=0$, equation (\ref{eq:shocktrajectory_soln}) yields $\tilde{y}  = 2/(1+\cos \tti)$, so
\beq \label{eq:SinExt-omegat}
\tti = \cos^{-1}\left(\frac{2}{\tilde{y}}-1\right).
\eeq
In this case $\Delta f = f(t_i)$; using $\sin[\cos^{-1}(x)] = \sqrt{1-x^2}$, 
\[
\Delta f = 2\frac{\sqrt{\tilde{y}(r)-1}}{\tilde{y}(r)}; 
\] 
and by equation (\ref{eq:dEwdr-dimensionless})), 
\beq 
\frac{d\tilde{E}_{\rm diss}}{d\tilde{y}} = \frac43 \frac{ (\tilde{y}-1)^{3/2}}{\tilde{y}^3}. 
\eeq 

The total energy deposited by a head or tail shock shock, $\tilde{E}_{\rm diss}(\infty)=\pi/2$, is only half of what is deposited at an internal shock, because external shocks interact with characteristics from only half a wave period.  However, external shocks expend this energy much more slowly.  For large $\tilde{y}$, external shocks obey $d\tilde{E}_{\rm diss}/d\tilde{y} \rightarrow (4/3)\tilde{y}^{-3/2}$, so the residual wave energy is 
\beq \label{eq:Ew-residual-ext}
\tilde{E}_w(\tilde{y}) \rightarrow \frac83 \tilde{y}^{-1/2},
\eeq 
in sharp contrast to internal shocks (eq.\ [\ref{eq:Ew-residual-int}]).  For this reason, internal shocks deliver more energy to radii of order the shock formation radius, while leading and trailing shocks have a much greater potential to create strong shocks in stellar atmospheres.  Figure~\ref{fig:ResidualEnergy-Sinusoid} compares the normalized residual energy $\tilde{E}_w(\tilde{y})$ between internal and head/tail shocks.

{\cdmrev Although we concentrate on purely sinusoidal wave trains, it is important to realize that the asymptotic properties of shock decay are not sensitive to the details of the waveform.  For head and tail shocks, for instance, the limit $f(t_{id})\rightarrow 2\tilde{y}^{-1/2}$ is a special case of the result
\begin{equation} \label{eq:HeadTailEqAarea} 
    f(y)^2 = \frac2y \int f(t_i) \,dt_i, 
\end{equation}
which can be obtained from the equal-area construction (see, for instance,  \citeauthor{whitham1974linear}), where the range of integration includes all those characteristics that will interact with the shock.   This construction is valid so long as the driving pulse is finite and significantly shorter than a preceding quiet period. (For it to hold at $y$ the quiet period must last a duration $y f(y)$.)  
} 

\begin{figure}[ht]
\includegraphics[width=0.5\textwidth]{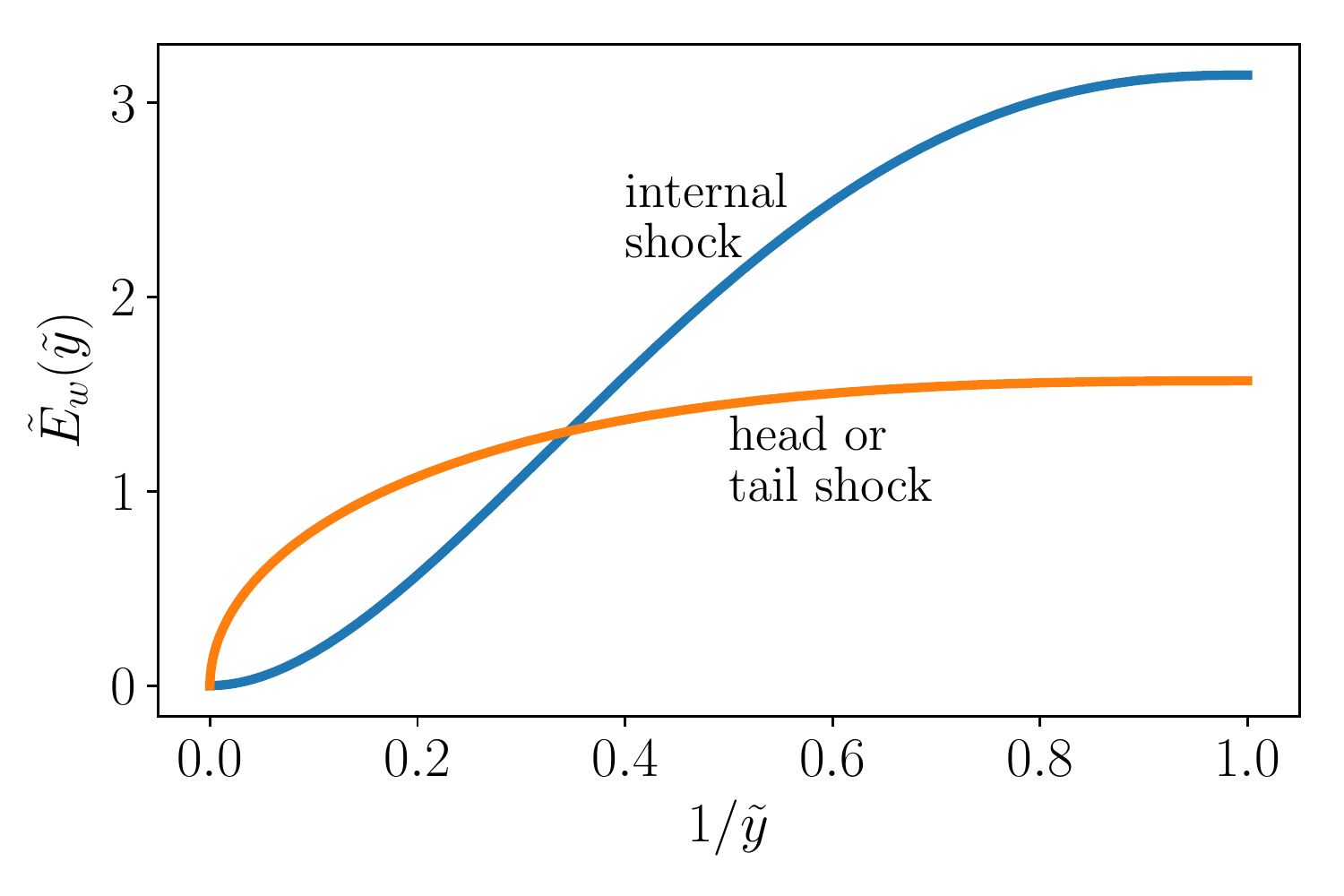}
\caption{Residual wave energy (normalized by $L_p/\omega$) available to drive internal and external (leading or trailing) shocks in a sinusoidal wave train, shown as a function of the inverse of the dimensionless evolution coordinate $\tilde{y}$.  Internal shocks, which deposit wave energy from an entire wave period, have twice the energy budget of an external shock.  However, external shocks preserve much more wave energy for late stages in their evolution.They therefore have a much greater capacity to drive strong shocks in stellar atmospheres. 
}
\label{fig:ResidualEnergy-Sinusoid}
\end{figure}  

} 
} 

\section{Hydrodynamic tests}
\label{sec:results}

To check our analytical results, we employ \texttt{FLASH4.6} \ (\citealt{2000ApJS..131..273F}) using its HLLC Riemann solver and 5th-order reconstruction. We resolve a spherical domain ($A = 4\pi r^2$ for $r_a<r<r_b$) with a uniform resolution and a static mesh, as we have found adaptive refinement boundaries to affect wave propagation.  We initialize either a power-law or polytropic hydrostatic structure, employing a static gravitational fields to balance forces in the background state.  A train of acoustic waves is introduced into the simulation domain by varying the fluid velocity, density, and pressure at the inner boundary. We fix the adiabatic index to $\gamma=5/3$ throughout. The simulation stops when the waves reach the (reflective) outer boundary.

\subsection{Power-Law Envelope}
\label{sec:powerlaw}

For our light, polytropic,  power-law atmosphere we consider $\rho_0\propto r^{-n}$ and adopt the Keplerian gravitiy of a central point mass, requiring $p_0\propto r^{-(n+1)}\propto \rho_0^{1+1/n}$, and $c_0\propto r^{-1/2}$.  We adopt $n=5/2$.   Our grid spans a factor of two in radius (5.6 in density), which we resolve with $1.4\times 10^5$ cells.  Into this we launch a train of sinusoidal waves ($v(r_a)=v_p \sin \omega t$ for $t>0$)  with $\omega  = 490 r_a/c_0(r_a)$, so the initial wavelength is $r_a/78$, or 1840 cells.  The initial amplitude is $v_p/c_0 = 1/100$.  At this amplitude and wavelength the linear high-frequency limit holds very well, and the numerical resolution is sufficient.  As we show in Figure~\ref{fig:sinusoid}, the simulation agrees very precisely with our analytical prediction for the radius of shock formation ($R_s=1.15 r_a$), as well as the development of the head shock and all subsequent internal shocks.   

The comparison is equally successful when we initialize a wave of only one sinusoidal cycle, which develops head and tail shocks but no internal shocks; see Figure~\ref{fig:sinusoid_single}.   (We have explored other waveforms, such as sawtooth and reverse-sawtooth waves.  These also show excellent agreement, in the linear regime, between analytical and numerical solutions.)

\begin{figure*}[ht]
\includegraphics[width=\textwidth]{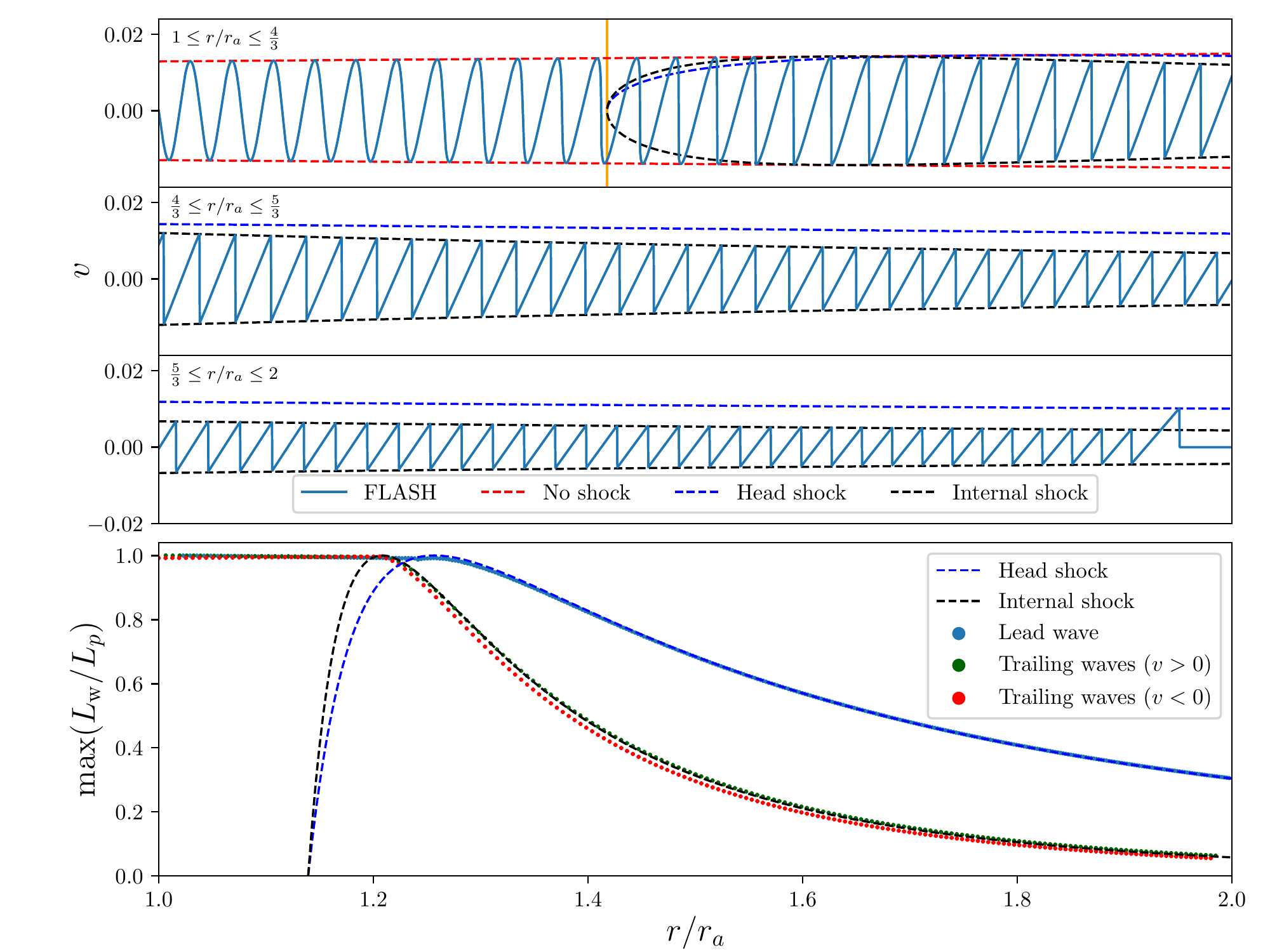}
\caption{Top: Velocity distribution of an outward traveling sinusoidal wave in an ambient medium with power-law structure (normalized by sound speed at the inner edge). The distribution is shown across the top three panels. Red dashed lines are the prediction for the maximum wave amplitude barring shock formation. Blue and black dashed lines are the predicted post-shock values for a forward and internal shock, respectively. Vertical yellow line is the predicted shock formation radius.  Bottom: Distribution of density, pressure, and relative entropy at several instances in time. The last panel shows a comparison of the predicted shock luminosity and measured normalized maximum wave luminosities.}
\label{fig:sinusoid}
\end{figure*}  

In the bottom panels of Figures\,\ref{fig:sinusoid} and \ref{fig:polytrope} we show the local maximum value of $L_w$ in units of the input peak value $L_p$, demonstrating that wave luminosity is conserved in the acoustic regime and decays as expected during shock evolution.

\begin{figure*}[ht]
\includegraphics[width=\textwidth]{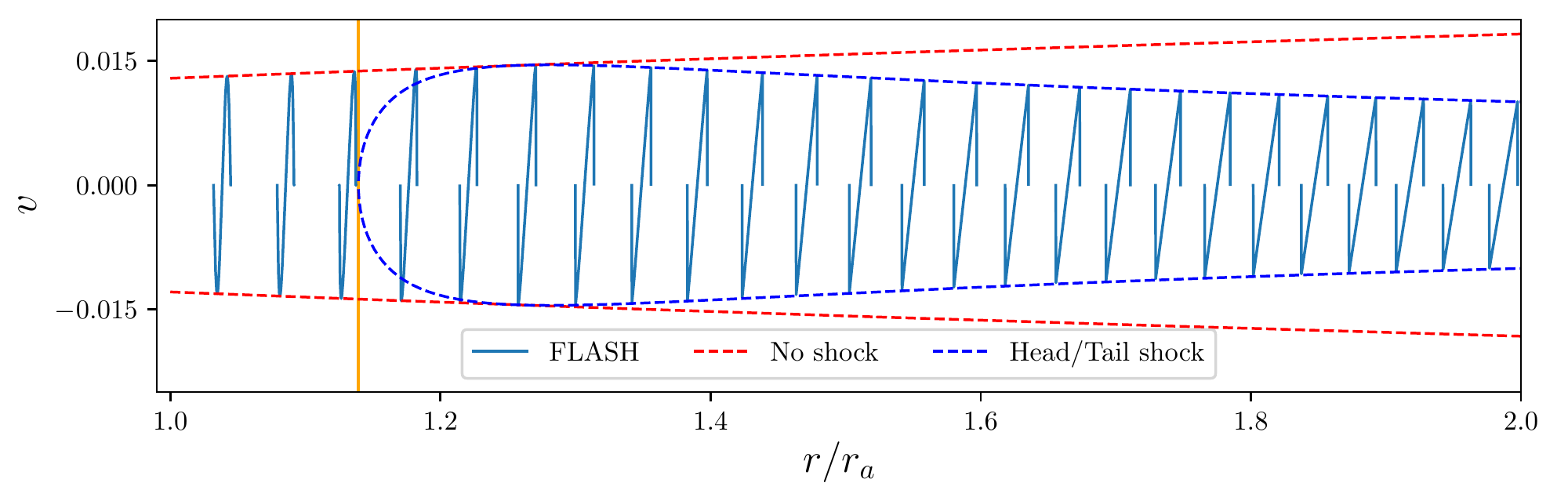}
\caption{Velocity distribution (normalized by sound speed at the inner edge) of a single sinusoidal wave in an ambient medium with power-law structure at several instances in time. The head (postive phase) and tail shocks (negative phase) evolve identically in linear heory. }
\label{fig:sinusoid_single}
\end{figure*}  



\begin{figure*}[ht]
\centering
\includegraphics[width=\textwidth]{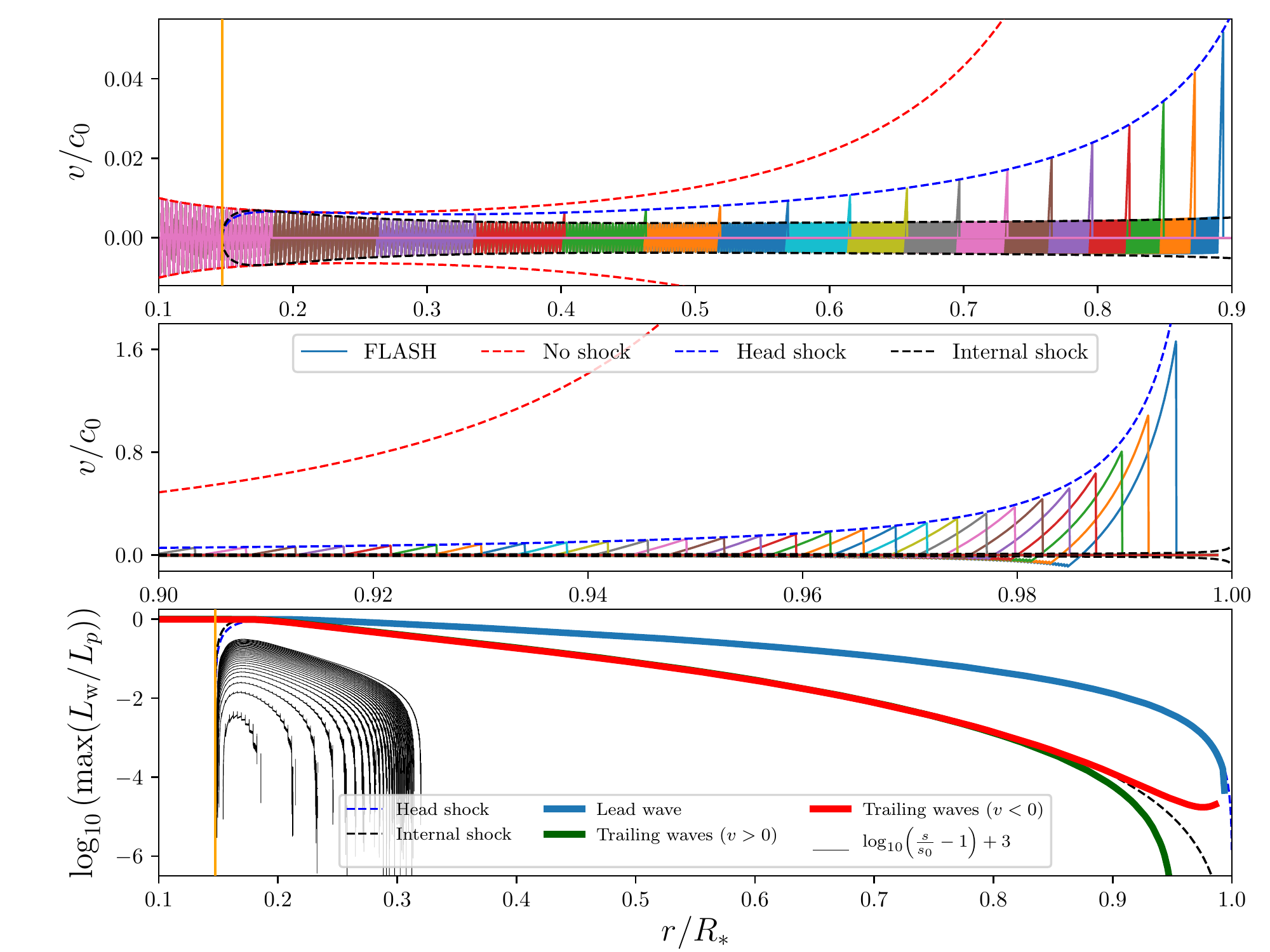}
\caption{Top and middle: Mach number distribution across an $n=2.5$ stellar polytrope excited by outward-traveling, small amplitude sinusoidal waves (i.e., $v/c_0=0.01$ at $r/r_a=1$). The distribution is divided into two spatial domains. Red dashed lines are the prediction for the maximum wave amplitude barring shock formation. Blue and black dashed lines are the predicted post-shock values given a forward and internal shock, respectively. Vertical yellow line is the predicted shock formation radius. Bottom: A comparison of the predicted shock luminosity and normalized maximum wave luminosities. Also shown are several instances of the relative entropy profile $s/s_0$ where $s=p/\rho^\gamma$. }
\label{fig:polytrope}
\end{figure*}

\subsection{Stellar Polytrope}

For a polytropic stellar model we adopt the self-gravitating Lane-Emden solution of radius $R_*$, in which $p_0\propto \rho_0^{1+1/n}$; again we choose  polytropic index $n=5/2$.    We arrange our grid to span radii $0.1R_*-0.999R_*$ with $1.5\times 10^5$ equally-spaced cells. Over this range $\rho_0$ varies by a factor of $2.6\times 10^8$ and $\Lwmax$ varies by a factor $2.9\times10^{11}$.   At the inner boundary we introduce sinusoidal waves of low amplitude ($v/c_0=10^{-2}$) and high frequency ($R_*/200$ in wavelength) so that each wave is initially resolved by approximately 800 cells.   This simulation is displayed in Figure~\ref{fig:polytrope}.  Again, a head shock and multiple internal shocks form at the predicted location ($R_s=0.147R_*$) and evolve as predicted in the linear regime.   

In contrast to the power-law atmosphere, however, the sharp drop in $\Lwmax$ near the stellar surface causes the head shock to become nonlinear.   The departure from linear theory is visible by $0.98 R_*$, where flow downstream of the head shock reaches $v/c_0 \sim 0.5$, and grows from there.  Linear theory remains remarkably accurate for flow Mach numbers of order unity.  New features, such as the presence of significant post-shock velocity, become visible as the head shock becomes strong. 

We note that the internal shocks remain weak as the head shock strengthens.  Indeed, the internal shocks continue to decelerate (because $V\approx c_0$ for a weak shock) as the head shock begins to accelerate; for this reason, a gap opens behind the advancing head shock.  (Tail shocks differ from head shocks in the nonlinear regime: they cannot strengthen and advance to the same degree.)   Although our simulation volume does not capture the region outside $R_*$, we can be confident that the head shock  develops sufficient strength to eject matter from the stellar surface.  

\section{Discussion}

%
%

\subsection{Shock decay and ejection: importance of the waveform} 

The most striking result of our analysis is the sharp difference in the decay properties of head and tail shocks, compared to internal shocks, in the simple model of a sinusoidal wave train composed of a finite number of cycles.   Consider the decay phase in which $f(t_{i,d})\rightarrow [\pi \tilde{y}^{-1}, 2\tilde{y}^{-1/2}] $ for [internal, head/tail] shocks, and let us define a characteristic  local propagation time 
\[ t_{\rm char}(r) \equiv Y(r) \Lwmax^{1/2}\] 
to eliminate $Y$.  Note that $t_{\rm char}$ need not be monotonic in $r$.   With this and our other definitions, the asymptotic shock strength is given by 
\begin{equation}  \label{eq:AsymptStrength_IntExt}
\frac{v_d^2}{c_0^2} = \frac{L_w(t_{i,d})}{\Lwmax} \rightarrow  \left\{ 
\begin{array}{lc}
\frac{\pi^2}{\omega^2 t_{\rm char}^2} & \mbox{internal~shock} \\  
\frac{4}{\omega t_{\rm char} }\frac{L_p^{1/2} }{\Lwmax(r)^{1/2}} & \mbox{head/tail~shock}
\end{array}
\right. 
\end{equation} 

Internal shocks, remarkably, do not depend in this limit on the initial wave amplitude; this implies a universal asymptotic shock heating behaviour that depends only on $\omega$ and local conditions, which we shall put to use in \S~\ref{SS:DissipationFormulae}.    Internal shocks dissipate wave energy relatively rapidly, but nevertheless strengthen in dimensionless terms in regions where $t_{\rm char}(r)$ declines outward. 

Head and tail shocks, in contrast, retain some sensitivity to the input luminosity, and strengthen once the combination $t_{\rm char}(r) \Lwmax(r)^{1/2}$ begins to decline.  Head and tail shocks evolve differently in the nonlinear regime;  head shocks accelerate more readily to become nascent blastwaves. 

Modelling the input wave as a pure sinusoidal wave train is highly idealized, so we consider the internal shocks and head or tail shocks as limiting cases of what would be observed in a real star.    We anticipate that a continuous but incoherent wave source will tend to resemble the internal shocks studied here, perhaps with a rare head- or tail-like shock arising by chance.  Transient events, on the other hand, may tend to resemble the head and tail shocks but also produce internal shocks in their ring-down phases.   We defer a full analysis to later work. 

\subsection{Dissipation formulae for wave luminosity} \label{SS:DissipationFormulae}

\cite{2018MNRAS.476.1853F} present a useful differential equation for the dissipation of wave energy at internal shocks, building on results from \citet{landau1959fluid}, \citet{1970SoPh...12..403U}, and \citet{1984oup..book.....M}.   We check this formula by examining the asymptotic behaviour of internal shocks in sinusoidal waves:  $\tilde{y}\rightarrow \pi/f$ for large $\tilde{y}$, from equation (\ref{eq:SinInt-y(ti)}).
Differentiating this and using definitions for $\tilde{y}$ and $f$ to express the result in terms of $L_w$, and then using $A \rho_0 dr = dm$ (defining the mass coordinate $m$), we find
\begin{equation}\label{eq:dLw-dm}
dL_w = - \frac{\gamma+1}{\pi} \omega c_0^2 \left(\frac{L_w}{\Lwmax}\right)^{3/2}dm . 
\end{equation} 
Note, $L_w$ here  means $L_{w}(t_{i,d})$ -- that is, it refers  to the the conditions at the shock.  The phase-averaged luminosity $\bar L_w$ is the more relevant  quantity.  Given that $L_w\propto f^2$, $\bar L_w = L_w(t_{i,d})/3$ in the asymptotic `N-wave' state; therefore 
\begin{equation}\label{eq:dLw-avg-dm} 
d\bar L_w = -\frac{\gamma+1}{\pi}\sqrt{3}  \omega c_0^2 \left(\frac{\bar L_w}{\Lwmax}\right)^{3/2} dm. 
\end{equation} 
The dissipation rate presented by \citet{2018MNRAS.476.1853F} agrees with this result, up to an inconsistency in their equation (17): on its left-hand side,  the symbol $L_w$ refers to $\bar L_w$, while on its right-hand side $L_w$ means $L_w(t_{id})=\bar L_w/3$.   Note also that equation (\ref{eq:dLw-avg-dm}) is accurate with our definition of $\Lwmax$, which is double the definition used by \citeauthor{2018MNRAS.476.1853F}.

Expression (\ref{eq:dLw-avg-dm}) is convenient in form, as it can be augmented to account for non-shock dissipation, as \citeauthor{2018MNRAS.476.1853F} have done.  However, it has a couple drawbacks. It applies only to the phase of asymptotic decay, so it cannot accurately describe the bulk of wave heating.  It can only be applied in the region $\tilde{y}>1$ where shocks exist; but even there, the initial value of $\bar L_w$ depends on what survives the early phase of shock evolution.   

For these reasons we recommend instead using the exact form of $\bar L_w(\tilde{y})$ that results from  recognizing that $\bar L_w$ is proportional to the residual wave energy $\tilde{E}_w$ as shown in Figure~\ref{fig:ResidualEnergy-Sinusoid}: 
\begin{equation} \label{eq:Exact-Lw-Decay-Internal}
\frac{\bar L_w(\tilde{y})}{\bar L_w(0)} = 1 -  \frac{\tilde{E}_{\rm diss}(\tti(\tilde{y}))}{\pi}
\end{equation} 
for $\tilde{y}>1$, where $\tilde{E}_{\rm diss}(\tti)$ is given by equation (\ref{eq:Ediss_cumulative-sin-internal}) and $\tti(\tilde{y}) = \sinc^{-1} (1/\tilde{y})$.   

Or, for an even more convenient alternative, an approximation based on the common asymptotic evolution of internal shocks in (eq.~\ref{eq:AsymptStrength_IntExt}, using $\bar L_w = L_w(t_{i,d})/3$): 
\begin{equation}  \label{eq:Appx-Lw-Decay-Internal}
\bar L_w(r) \simeq \left[  \bar L_w(0)^{-k} + \left( \frac{\pi^2 \Lwmax(r)}{3 \omega^2 t_{\rm char}(r)^2}\right)^{-k}\right]^{-1/k} 
\end{equation} 
which respects both limits, and remains within $10\%$ for $k=0.64$. 

Both the exact and approximate expressions (eqs.~\ref{eq:Exact-Lw-Decay-Internal} and \ref{eq:Appx-Lw-Decay-Internal}) assume the input wave is a pure uninterrupted sinusoid that remains in the linear, high frequency regime, so its shock dissipation can be predicted by the results of \S~\ref{SSS:InternalShocks}.  For a more sophisticated approach one must start from a more realistic input waveform.  These expressions also neglect other types of dissipation, such as radiation diffusion; we \citep{2017ApJ...841....9R} have found this to be valid when $\bar L_w$ exceeds the  radiation diffusion luminosity of the background state. 

{\cdmrev 
\subsection{Surface ejection by head shocks: weak limit} \label{SS:mass-ejection}
We wish to make a rough estimate for the mass ejected by a head shock that strengthens in the outer stellar envelope, in the weak-wave limit for which this occurs very close to the stellar surface ($z \equiv (R_*-r_0)/R_* \ll 1$, where $R_*$ is the stellar radius).  We begin with ejection by a weak acoustic pulse, in which shock formation also occurs in the outer envelope, and then consider the case of a central point explosion in the next subsection. 

Our analysis here is only asymptotic, however, as we neglect radiation diffusion, which would truncate the acceleration toward $v_{\rm esc *}$, the stellar escape velocity, for sufficiently low wave energies (Linial et al.\ 2020, in prep.).  We also require that the wave be sufficiently radial that lateral pressure gradients do not alter the character of the flow \citep{MatznerLevinRo13_Oblique,Salbi14_Oblique,LinialSari19}.

The outer polytropic scalings are $A=4\pi R_*^2$, $\rho_0 = \rho_{0h} z^n$, and $c_0 = c_{0h}z^{1/2}$, where the `h' subscript indicates the outer normalization and $z = (R_*-r_0)/R_*$ is the relative depth.  From these we derive the external mass $m_h z^{n+1}$ (where $m_h=4\pi R_*^3 \rho_{0h}/(n+1)$),  maximum luminosity  $\Lwmax = L_{{\rm max},h} z^{n+3/2}$ (where $ L_{{\rm max},h} = 4\pi R_*^2 \rho_{0,h}c_{0,h}^3$), and  characteristic time $t_{\rm char} = t_{{\rm char},h} z^{1/2}$ (where $t_{{\rm char},h} =  (\gamma+1)(2n+1)R_*/8 c_{0,h})$). 

For a reasonable approximation to the ejected mass, we first identify the value of $z=z_S$ at which the shock becomes strong, in the sense that  $v_d(z_S)=c_0(z_S)$, using equation (\ref{eq:AsymptStrength_IntExt}) to extrapolate the head shock out of the weak shock regime. We determine the corresponding shock velocity $v_s(z_S) =\left( \gamma+1 + \sqrt{17+\gamma(2+\gamma})\right)c_0(z_S)/4$.  We then apply the self-similar scalings for a strong planar shock, $ v_s\propto  \rho_0^{-\beta}$ \citep{sakurai60} until the ejection condition $f_{\rm sph} C v_s = v_{\rm esc *}$ is met.  Here $v_{\rm esc *}$ is the stellar escape velocity, $\beta \simeq 0.19$ is the shock acceleration index, and $C\sim 2$ is the post-shock acceleration factor for purely planar, non-relativistic flow \citep{sakurai60,RoMatzner13_shocks}.  The factor $f_{\rm sph}$ accounts for the difference between spherical and planar flow, which matters for finite but small $E_{\rm in}$ because of the slow nature of post-shock acceleration \citep{LitvinovaNadezhin90}; for non-gravitating flow $f_{\rm sph}\simeq (1-\{0.51,0.34\}z^{1/3})$ for $n=\{\frac32,3\}$ \citep{1999ApJ...510..379M}. 

We must also allow for the possibility of wave reflection, in the case that the high-frequency limit does not hold throughout its transition to a strong, accelerating shock. (Reflection is not an issue in the strong regime, as \citeauthor{sakurai60}'s solution accounts for it self-consistently.)    The high-frequency limit requires $\omega z R_*/c_0 \gg1$; evaluating this at the weak-to-strong transition, 
\begin{equation}
\omega \frac{z_S R_*}{c_0(z_S)} =
\left[\frac{L_p}{L_{{\rm max},h}} \left(\frac{t_{{\rm char},h} c_{0h}}{4R_*}\right)^2 \left(\frac{c_{0h}}{\omega R_*}\right)^{3+2n}   \right]^{\frac1{5+2n}}. 
\end{equation}  

As an analysis of wave reflection is beyond the scope of this paper, we capture it using a velocity transmission prefactor $f_{\rm tm}\leq 1$, where $f_{\rm tm} = 1$ in the absence of reflection.  Then
\begin{eqnarray} \label{eq:EjectedMass-Wave}
\frac{m_{\rm ej} } {m_h } =
\left[ f_{\rm tm} f_{\rm sph} K \frac{c_{0h}}{v_{\rm esc *}}
\left(\frac{L_p}{L_{{\rm max},h}} \frac{16}{\omega^2 t_{\rm char,h}^2}\right)^\frac{1+2\beta n}{5+2n} \right]^{\gamma_p/\beta}\\
\longrightarrow \left(\frac{3.47 f_{\rm tm}f_{\rm sph}\,c_{0h}}{v_{\rm esc *}}
\right)^{7.18}\left(\frac{L_p}{L_{{\rm max},h}} \frac{1}{\omega^2 t_{{\rm char},h}^2}\right)^{1.38}\nonumber 
\end{eqnarray}
The coefficient $K\simeq C v_s(z_S)/c_0(z_S)$ depends on our approximate matching between weak and strong limits. 
The evaluation on the second line of (\ref{eq:EjectedMass-Wave}) is for radiation-dominated flow in an $n=3$ polytrope ($\gamma=\gamma_p=4/3$).

\subsection{Mass ejection from a weak central explosion}\label{SS:Ejection-Explosion}

Our results can also be applied to the similar problem of mass ejection caused by a  explosion of small but finite strength in the stellar centre, which has been studied by \citet{WymanChernoffWasserman04} and Linial, Sari, \& Fuller (2020, in prep.) for polytropes, and by \citet{2010MNRAS.405.2113D}, \citet{Owocki19_eruptions}, and \citet{KuriyamaShigeyama20_outbursts} for realistic stellar models.  We consider polytropes for simplicity.  The explosion is strong until it decelerates to about $c_0$; for a complete polytrope, this occurs on a time $t_{\rm dec}\simeq
[2/(5c_{0c})]^{5/3} (E_{\rm in}/\rho_{0c}) ^{1/3}$ for an explosion of energy $E_{\rm in}$ that is small compared to the binding energy; subscript $c$ indicates conditions at the center, and `dec' refers to the conditions at the deceleration radius.   

While there is no shock-free phase in this problem, we can nevertheless, for a good approximation, adopt a head shock model in which the shock forms at the deceleration radius ($r_s \simeq r_{\rm dec}$); the energy in the wave pulse is $\sim E_{\rm in}/2$ (accounting for heat deposited within $r_{\rm dec}$); and the pulse duration, or  wave half-period, is $\pi/\omega \simeq t_{\rm dec}$.  Correspondingly, $L_p\simeq E_{\rm in}/t_{\rm dec}$.   

We require the normalized coordinate $\tilde{y}(r)=Y(r)/Y_s =Y(r)/Y(r_{\rm dec})$.  For the denominator, we note that $Y(r)$ is almost constant in a uniform medium:  $Y(r_{\rm dec})=\Lambda (\gamma+1)(4\pi c_{0c}\rho_{0c}^5)^{1/2}/2$; we drop $\Lambda$, which is a logarithm of $r_{\rm dec}$.   For the numerator, $Y(z)=2[(\gamma+1)/(2n+1)]z R_*/[c_0(z) \Lwmax(z)^{1/2}]$ in the sub-surface region ($z\ll1)$.  

We now apply the asymptotic head shock formula $v_d^2/c_0^2 = (4/\tilde{y}) L_p/\Lwmax$ to identify $z_S$ as the depth at which the shock regains its strength ($v_d=c_0$), and proceed to estimate $m_{\rm ej}$ as before.   The full expression is too cumbersome to write, but we note that the ejected mass scales $\propto (f_{\rm tm} f_{\rm sph})^{\gamma_p/\beta} E_{\rm in}^\alpha$, where
\begin{equation}
    \alpha = \frac{4\gamma_p (1+2\beta n)}{3\beta(5+2n)}.
\end{equation}
The coefficient can be determined from the polytropic structure, which fixes  $(\rho_{0h}/\rho_{0c})^{1/n} = (c_{0h}/c_{0c})^2 = (\xi \theta')_{\xi_n}$ where $\theta(\xi)$ is the Lane-Emden function with outer coordinate $\xi_n$.  For the case $n=3/2$ and $\gamma=5/3$,  we obtain 
\begin{equation} \label{eq:EjectedMass-Expln}
\frac{m_{\rm ej}} {M_*} \simeq \left(2.23f_{\rm tm}f_{\rm sph}\right)^{7.52} \left(\frac{E_{\rm in}}{E_{\rm bind}}\right)^{2.09}
\end{equation}
where $E_{\rm bind}$ is the negative total energy of the polytrope. This is consistent with the preliminary results of Linial and collaborators (Itai Linial, private communication, Oct.~2020) if  $K f_{\rm tm}f_{\rm sph}\propto  (E_{\rm in}/E_{\rm bind})^{0.12}$.  

We suspect that the threshold energy observed by \citet{WymanChernoffWasserman04} is a consequence of  finite numerical resolution in that work.  However, radiation diffusion sets a true lower limit for $m_{\rm ej}$. For analysis of this point, see \citet{LinialSariFuller20}.

}

\subsection{Failed Supernova Shocks}
A `failed supernova' is a collapsing massive star in which the stalled protoneutron star bounce shock fails to revive and eject the envelope. Instead, the protoneutron star is expected to collapse into a black hole since the total accreted mass will exceed the Tolman-Oppenheimer-Volkoff limit. The nearly instantaneous liberation of $\sim\!0.3\,\msun$ in neutrinos from the formation of the protoneutron star causes the over-pressurized envelope to accelerate outward at all radii, at least initially. The expansion launches an outward-propagating spherical acoustic pulse that steepens into a weak shock. In the wake of the shock is an expanding rarefaction wave that informs the exterior envelope of a collapse in the stellar core. 

Red and yellow supergiant stars are well-approximated by the light polytropic envelope structure described in \S~\ref{sec:powerlaw}, with index $n=2.5$, over several decades of $r$.   For all positive values of $n$, there are at least two self-similar solutions to the flow structure: one in which the weak shock becomes strong and is described by the Sedov-Taylor solutions, and another where its strength vanishes and it merges with the leading edge of the rarefaction wave.   For the range $2<n<3.5$, \cite{cqr1} discovered a third set of intermediate self-similar solutions involving weak shocks of a finite strength. However, \cite{cqr2} show by linear analysis that these self-similar weak shock solutions are linearly unstable to perturbations in the shock strength, and  \cite{cqr3} show these shocks asymptotically approach the rarefaction and Sedov-Taylor self-similar solutions.

Why do these intermediate solutions only exist for $2<n<3.5$? The lower limit presumably derives from the fact that, for $n<2$ the Sedov-Taylor blastwave weakens over time, as it implies $V/c_0 \propto t^{(n-2)/(5-n)}$.    For $n\ge3.5$, \cite{cqr3} found that all shocks strengthen, evolving toward the Sedov-Taylor limit. Using our solutions from \S\ref{sec:particular_solutions}, we find that a head shock obeys $v_d/c_0 \propto r\sqrt{A \rho_0 c_0}\propto r^{(n-7/2)/4}$: it strengthens over time when $n>3.5$.  In these conditions, a weak shock tends to outrun outward characteristics, causally disconnecting itself from the collapsing interior.

\section{Conclusion} \label{sec:conclusion}

A main conclusion of this work is that shocks created by radially-propagating sound waves -- at least in the relevant limit of high frequencies and small amplitudes -- can be captured using a straightforward extension of the classical equal-area method, either in its integral or differential form.  The future shock evolution can effectively be read from the initial waveform and the stellar profile of density and sound speed.  The method leads to improved expressions for the shock strength and phase-averaged luminosity that should find application in the modelling of pre-supernova outbursts.   

Any detailed prediction for shock evolution, however, requires some knowledge of how the wave is driven.  This is very clear for the case we study, in which the input waveform is a sinusoidal function that abruptly starts and stops at its wave nodes.  Such waveforms generate internal shocks, which dissipate rapidly after shock formation, as well as head and tail shocks (if they begin or end in a compression) that decay slowly, and that preserve significant energy even to the stellar surface -- as in the example shown in Figure~\ref{fig:polytrope}.     The distinct potential of head shocks to drive shock-driven outbursts is a novel aspect of our analysis. 

We highlight several avenues for future study: 

\noindent 1.\ Examination of strengthening shocks.   Weak shocks, especially head shocks, can evolve to become strong in sufficiently stratified regions like outer stellar envelopes.  There already are approximate theories that handle this transition \citep[e.g.,][]{BrinkleyKirkwood47,ro2017launching} but  there may exist a smooth interpolation into the strong-shock regime, like the one used by \citet{2001ApJ...551..946T} to describe trans-relativistic explosions, which would improve on the estimate we give in   \S\,\ref{SS:mass-ejection}.  {\cdmrev One should also investigate the potential for some of the wave energy to be reflected before the transition is complete, as well as spherical effects in post-shock acceleration.}

\noindent 2.\ Analysis of more general waveforms, such as Gaussian random noise with a specified power spectrum.  We anticipate that the dissipation profile, as well as the rate at which head- and tail-like shocks are generated at random, will depend on the effective bandwidth of the input spectrum. 

\noindent 3.\ Adaptation to moving backgrounds.  The current analysis is limited to hydrostatic environments, but significant motions could be present either in the initial state or because of previous shocks.  

\noindent 4.\ Characterization of the driving source.  For transient events such as the pulsational pair instability this would involve a semi-analytical analysis, or  direct numerical simulation until the emerging waves are well into the high-frequency regime.  For vigorous convection this might be accomplished as discussed above. 

\noindent 5.\ Extension to multipole emission and multi-dimensional propagation.  Although emission from the core will tend to be of low or zero multipole order, some driving sources, like shell convection or compact common-envelope in-spiral, are intrinsically multipolar.   \cite{lighthill78} discusses how geometrical acoustics can be used to adapt one-dimensional analyses to the case of non-radial propagation. 

\noindent 6.\ Extension to moving flows. This will be necessary, for instance, to test \citet{LeungFuller20}'s scenario in which the shock dissipation of high-frequency waves creates a low-frequency wave, which subsequently shocks. 

\noindent 7.\ Further studies of shock-driven ejection.   Models like those of \citet{2010MNRAS.405.2113D}, \citet{2017MNRAS.470.1642F}, and \citet{2018MNRAS.476.1853F} can be connected to the specifics of the driving waveform using the tools developed here.

\section*{Acknowledgements}

{\cdmrev We thank Itai Linial, Re'em Sari, and Jim Fuller for sharing their work prior to publication, and Jim Fuller for clarifying questions.}  This work was funded by an NSERC Discovery Grant (CDM) and by the  Gordon and Betty Moore Foundation through Grant GBMF5076 (SR).  

\software{ \texttt{FLASH4.6} \ (\citealt{2000ApJS..131..273F})  }

{\em Greenhouse gas emissions:} We estimate $100$\,kg CO$_2$ equivalent from  simulations presented here, on the basis of $\sim 400$ kWh of Berkeley, CA electricity. 

\bibliography{bib}

\begin{thebibliography}{}
\expandafter\ifx\csname natexlab\endcsname\relax\def\natexlab#1{#1}\fi

\bibitem[{{Abbott} {et~al.}(2020){Abbott}, {Abbott}, {Abraham}, \&
  {Acernese}}]{LIGO2020_GW190521}
{Abbott}, R., {Abbott}, T.~D., {Abraham}, S., \& {Acernese}, F. 2020, arXiv
  e-prints, arXiv:2009.01190

\bibitem[{{Brinkley} \& {Kirkwood}(1947)}]{BrinkleyKirkwood47}
{Brinkley}, S.~R., \& {Kirkwood}, J.~G. 1947, Physical Review, 71, 606

\bibitem[{Chugai {et~al.}(2004)Chugai, Blinnikov, Cumming, Lundqvist,
  Bragaglia, Filippenko, Leonard, Matheson, \& Sollerman}]{Chugai04}
Chugai, N.~N., Blinnikov, S.~I., Cumming, R.~J., {et~al.} 2004, Monthly Notices
  of the Royal Astronomical Society, 352, 1213

\bibitem[{{Coughlin} {et~al.}(2018){Coughlin}, {Quataert}, \& {Ro}}]{cqr1}
{Coughlin}, E.~R., {Quataert}, E., \& {Ro}, S. 2018, \apj, 863, 158

\bibitem[{{Coughlin} {et~al.}(2019){Coughlin}, {Ro}, \& {Quataert}}]{cqr2}
{Coughlin}, E.~R., {Ro}, S., \& {Quataert}, E. 2019, arXiv e-prints,
  arXiv:1901.04487

\bibitem[{{Courant} \& {Friedrichs}(1948)}]{1948sfsw.book.....C}
{Courant}, R., \& {Friedrichs}, K.~O. 1948, {Supersonic flow and shock waves}

\bibitem[{{Dessart} {et~al.}(2010){Dessart}, {Livne}, \&
  {Waldman}}]{2010MNRAS.405.2113D}
{Dessart}, L., {Livne}, E., \& {Waldman}, R. 2010, \mnras, 405, 2113

\bibitem[{{Farmer} {et~al.}(2019){Farmer}, {Renzo}, {de Mink}, {Marchant}, \&
  {Justham}}]{Farmer19_PPI}
{Farmer}, R., {Renzo}, M., {de Mink}, S.~E., {Marchant}, P., \& {Justham}, S.
  2019, \apj, 887, 53

\bibitem[{{Foley} {et~al.}(2007){Foley}, {Smith}, {Ganeshalingam}, {Li},
  {Chornock}, \& {Filippenko}}]{2007ApJ...657L.105F}
{Foley}, R.~J., {Smith}, N., {Ganeshalingam}, M., {et~al.} 2007, \apjl, 657,
  L105

\bibitem[{Friedrichs(1948)}]{friedrichs48}
Friedrichs, K.~O. 1948, Communications on Pure and Applied Mathematics, 1, 211

\bibitem[{{Fryxell} {et~al.}(2000){Fryxell}, {Olson}, {Ricker}, {Timmes},
  {Zingale}, {Lamb}, {MacNeice}, {Rosner}, {Truran}, \&
  {Tufo}}]{2000ApJS..131..273F}
{Fryxell}, B., {Olson}, K., {Ricker}, P., {et~al.} 2000, \apjs, 131, 273

\bibitem[{{Fuller}(2017)}]{2017MNRAS.470.1642F}
{Fuller}, J. 2017, \mnras, 470, 1642

\bibitem[{{Fuller} \& {Ro}(2018)}]{2018MNRAS.476.1853F}
{Fuller}, J., \& {Ro}, S. 2018, \mnras, 476, 1853

\bibitem[{{Gal-Yam} {et~al.}(2014){Gal-Yam}, {Arcavi}, {Ofek}, {Ben-Ami},
  {Cenko}, {Kasliwal}, {Cao}, {Yaron}, {Tal}, {Silverman}, {Horesh}, {De Cia},
  {Taddia}, {Sollerman}, {Perley}, {Vreeswijk}, {Kulkarni}, {Nugent},
  {Filippenko}, \& {Wheeler}}]{2014Natur.509..471G}
{Gal-Yam}, A., {Arcavi}, I., {Ofek}, E.~O., {et~al.} 2014, \nat, 509, 471

\bibitem[{{Groh}(2014)}]{2014A&A...572L..11G}
{Groh}, J.~H. 2014, \aap, 572, L11

\bibitem[{{Hosseinzadeh} {et~al.}(2017){Hosseinzadeh}, {Arcavi}, {Valenti},
  {McCully}, {Howell}, {Johansson}, {Sollerman}, {Pastorello}, {Benetti},
  {Cao}, {Cenko}, {Clubb}, {Corsi}, {Duggan}, {Elias-Rosa}, {Filippenko},
  {Fox}, {Fremling}, {Horesh}, {Karamehmetoglu}, {Kasliwal}, {Marion}, {Ofek},
  {Sand}, {Taddia}, {Zheng}, {Fraser}, {Gal-Yam}, {Inserra}, {Laher}, {Masci},
  {Rebbapragada}, {Smartt}, {Smith}, {Sullivan}, {Surace}, \&
  {Wo{\'z}niak}}]{2017ApJ...836..158H}
{Hosseinzadeh}, G., {Arcavi}, I., {Valenti}, S., {et~al.} 2017, \apj, 836, 158

\bibitem[{{Khazov} {et~al.}(2016){Khazov}, {Yaron}, {Gal-Yam}, {Manulis},
  {Rubin}, {Kulkarni}, {Arcavi}, {Kasliwal}, {Ofek}, {Cao}, {Perley},
  {Sollerman}, {Horesh}, {Sullivan}, {Filippenko}, {Nugent}, {Howell}, {Cenko},
  {Silverman}, {Ebeling}, {Taddia}, {Johansson}, {Laher}, {Surace},
  {Rebbapragada}, {Wozniak}, \& {Matheson}}]{2016ApJ...818....3K}
{Khazov}, D., {Yaron}, O., {Gal-Yam}, A., {et~al.} 2016, \apj, 818, 3

\bibitem[{{Kiewe} {et~al.}(2012){Kiewe}, {Gal-Yam}, {Arcavi}, {Leonard},
  {Emilio Enriquez}, {Cenko}, {Fox}, {Moon}, {Sand}, {Soderberg}, \&
  {CCCP}}]{2012ApJ...744...10K}
{Kiewe}, M., {Gal-Yam}, A., {Arcavi}, I., {et~al.} 2012, \apj, 744, 10

\bibitem[{{Kuriyama} \& {Shigeyama}(2020)}]{KuriyamaShigeyama20_outbursts}
{Kuriyama}, N., \& {Shigeyama}, T. 2020, \aap, 635, A127

\bibitem[{Landau \& Lifshitz(1959)}]{landau1959fluid}
Landau, L., \& Lifshitz, E. 1959, Fluid mechanics, A-W series in advanced
  physics (Pergamon Press)

\bibitem[{Landau(1945)}]{landau1945shock}
Landau, L.~D. 1945, J. Phys. USSR, 9, 496

\bibitem[{{Leung} \& {Fuller}(2020)}]{LeungFuller20}
{Leung}, S.-C., \& {Fuller}, J. 2020, \apj, 900, 99

\bibitem[{{Leung} {et~al.}(2019){Leung}, {Nomoto}, \&
  {Blinnikov}}]{LeungNomotoBlinnikov19}
{Leung}, S.-C., {Nomoto}, K., \& {Blinnikov}, S. 2019, arXiv e-prints,
  arXiv:1901.11136

\bibitem[{{Lighthill}(1978)}]{lighthill78}
{Lighthill}, J. 1978, {Waves in fluids} (Cambridge University Press)

\bibitem[{{Lin} \& {Szeri}(2001)}]{2001JFM...431..161L}
{Lin}, H., \& {Szeri}, A.~J. 2001, Journal of Fluid Mechanics, 431, 161

\bibitem[{{Linial} \& {Sari}(2019)}]{LinialSari19}
{Linial}, I., \& {Sari}, R. 2019, Physics of Fluids, 31, 097102

\bibitem[{{Linial} {et~al.}(2020){Linial}, {Sari}, \&
  {Fuller}}]{LinialSariFuller20}
{Linial}, I., {Sari}, R., \& {Fuller}, J. 2020, in prep.

\bibitem[{{Litvinova} \& {Nadezhin}(1990)}]{LitvinovaNadezhin90}
{Litvinova}, I.~Y., \& {Nadezhin}, D.~K. 1990, Soviet Astronomy Letters, 16, 29

\bibitem[{{Margutti} {et~al.}(2014){Margutti}, {Milisavljevic}, {Soderberg},
  {Chornock}, {Zauderer}, {Murase}, {Guidorzi}, {Sanders}, {Kuin}, {Fransson},
  {Levesque}, {Chandra}, {Berger}, {Bianco}, {Brown}, {Challis},
  {Chatzopoulos}, {Cheung}, {Choi}, {Chomiuk}, {Chugai}, {Contreras}, {Drout},
  {Fesen}, {Foley}, {Fong}, {Friedman}, {Gall}, {Gehrels}, {Hjorth}, {Hsiao},
  {Kirshner}, {Im}, {Leloudas}, {Lunnan}, {Marion}, {Martin}, {Morrell},
  {Neugent}, {Omodei}, {Phillips}, {Rest}, {Silverman}, {Strader},
  {Stritzinger}, {Szalai}, {Utterback}, {Vinko}, {Wheeler}, {Arnett},
  {Campana}, {Chevalier}, {Ginsburg}, {Kamble}, {Roming}, {Pritchard}, \&
  {Stringfellow}}]{2014ApJ...780...21M}
{Margutti}, R., {Milisavljevic}, D., {Soderberg}, A.~M., {et~al.} 2014, \apj,
  780, 21

\bibitem[{{Margutti} {et~al.}(2017){Margutti}, {Kamble}, {Milisavljevic},
  {Zapartas}, {de Mink}, {Drout}, {Chornock}, {Risaliti}, {Zauderer},
  {Bietenholz}, {Cantiello}, {Chakraborti}, {Chomiuk}, {Fong}, {Grefenstette},
  {Guidorzi}, {Kirshner}, {Parrent}, {Patnaude}, {Soderberg}, {Gehrels}, \&
  {Harrison}}]{2017ApJ...835..140M}
{Margutti}, R., {Kamble}, A., {Milisavljevic}, D., {et~al.} 2017, \apj, 835,
  140

\bibitem[{{Matzner} {et~al.}(2013){Matzner}, {Levin}, \&
  {Ro}}]{MatznerLevinRo13_Oblique}
{Matzner}, C.~D., {Levin}, Y., \& {Ro}, S. 2013, \apj, 779, 60

\bibitem[{{Matzner} \& {McKee}(1999)}]{1999ApJ...510..379M}
{Matzner}, C.~D., \& {McKee}, C.~F. 1999, \apj, 510, 379

\bibitem[{{Matzner} \& {Ro}(2020)}]{2019jfm}
{Matzner}, C.~D., \& {Ro}, S. 2020, Journal~of~Fluid~Mechanics, submitted,
  arXiv:2011.07180

\bibitem[{{Mihalas} \& {Mihalas}(1984)}]{1984oup..book.....M}
{Mihalas}, D., \& {Mihalas}, B.~W. 1984, {Foundations of radiation
  hydrodynamics}

\bibitem[{{Milisavljevic} {et~al.}(2015){Milisavljevic}, {Margutti}, {Kamble},
  {Patnaude}, {Raymond}, {Eldridge}, {Fong}, {Bietenholz}, {Challis},
  {Chornock}, {Drout}, {Fransson}, {Fesen}, {Grindlay}, {Kirshner}, {Lunnan},
  {Mackey}, {Miller}, {Parrent}, {Sanders}, {Soderberg}, \&
  {Zauderer}}]{2015ApJ...815..120M}
{Milisavljevic}, D., {Margutti}, R., {Kamble}, A., {et~al.} 2015, \apj, 815,
  120

\bibitem[{{Moriya} {et~al.}(2014){Moriya}, {Maeda}, {Taddia}, {Sollerman},
  {Blinnikov}, \& {Sorokina}}]{2014MNRAS.439.2917M}
{Moriya}, T.~J., {Maeda}, K., {Taddia}, F., {et~al.} 2014, \mnras, 439, 2917

\bibitem[{{Ofek} {et~al.}(2016){Ofek}, {Cenko}, {Shaviv}, {Duggan},
  {Strotjohann}, {Rubin}, {Kulkarni}, {Gal-Yam}, {Sullivan}, {Cao}, {Nugent},
  {Kasliwal}, {Sollerman}, {Fransson}, {Filippenko}, {Perley}, {Yaron}, \&
  {Laher}}]{Ofek16_2015bh}
{Ofek}, E.~O., {Cenko}, S.~B., {Shaviv}, N.~J., {et~al.} 2016, \apj, 824, 6

\bibitem[{{Owocki} {et~al.}(2019){Owocki}, {Hirai}, {Podsiadlowski}, \&
  {Schneider}}]{Owocki19_eruptions}
{Owocki}, S.~P., {Hirai}, R., {Podsiadlowski}, P., \& {Schneider}, F. R.~N.
  2019, \mnras, 485, 988

\bibitem[{{Pastorello} {et~al.}(2007){Pastorello}, {Smartt}, {Mattila},
  {Eldridge}, {Young}, {Itagaki}, {Yamaoka}, {Navasardyan}, {Valenti}, {Patat},
  {Agnoletto}, {Augusteijn}, {Benetti}, {Cappellaro}, {Boles}, {Bonnet-Bidaud},
  {Botticella}, {Bufano}, {Cao}, {Deng}, {Dennefeld}, {Elias-Rosa},
  {Harutyunyan}, {Keenan}, {Iijima}, {Lorenzi}, {Mazzali}, {Meng}, {Nakano},
  {Nielsen}, {Smoker}, {Stanishev}, {Turatto}, {Xu}, \&
  {Zampieri}}]{2007Natur.447..829P}
{Pastorello}, A., {Smartt}, S.~J., {Mattila}, S., {et~al.} 2007, \nat, 447, 829

\bibitem[{{Pastorello} {et~al.}(2008{\natexlab{a}}){Pastorello}, {Mattila},
  {Zampieri}, {Della Valle}, {Smartt}, {Valenti}, {Agnoletto}, {Benetti},
  {Benn}, {Branch}, {Cappellaro}, {Dennefeld}, {Eldridge}, {Gal-Yam},
  {Harutyunyan}, {Hunter}, {Kjeldsen}, {Lipkin}, {Mazzali}, {Milne},
  {Navasardyan}, {Ofek}, {Pian}, {Shemmer}, {Spiro}, {Stathakis},
  {Taubenberger}, {Turatto}, \& {Yamaoka}}]{2008MNRAS.389..113P}
{Pastorello}, A., {Mattila}, S., {Zampieri}, L., {et~al.} 2008{\natexlab{a}},
  \mnras, 389, 113

\bibitem[{{Pastorello} {et~al.}(2008{\natexlab{b}}){Pastorello}, {Mattila},
  {Zampieri}, {Della Valle}, {Smartt}, {Valenti}, {Agnoletto}, {Benetti},
  {Benn}, {Branch}, {Cappellaro}, {Dennefeld}, {Eldridge}, {Gal-Yam},
  {Harutyunyan}, {Hunter}, {Kjeldsen}, {Lipkin}, {Mazzali}, {Milne},
  {Navasardyan}, {Ofek}, {Pian}, {Shemmer}, {Spiro}, {Stathakis},
  {Taubenberger}, {Turatto}, \& {Yamaoka}}]{Pastorello08a-Herich}
---. 2008{\natexlab{b}}, \mnras, 389, 113

\bibitem[{{Pastorello} {et~al.}(2008{\natexlab{c}}){Pastorello}, {Quimby},
  {Smartt}, {Mattila}, {Navasardyan}, {Crockett}, {Elias-Rosa}, {Mondol},
  {Wheeler}, \& {Young}}]{2008MNRAS.389..131P}
{Pastorello}, A., {Quimby}, R.~M., {Smartt}, S.~J., {et~al.}
  2008{\natexlab{c}}, \mnras, 389, 131

\bibitem[{{Pastorello} {et~al.}(2013){Pastorello}, {Cappellaro}, {Inserra},
  {Smartt}, {Pignata}, {Benetti}, {Valenti}, {Fraser}, {Tak{\'a}ts}, {Benitez},
  {Botticella}, {Brimacombe}, {Bufano}, {Cellier-Holzem}, {Costado}, {Cupani},
  {Curtis}, {Elias-Rosa}, {Ergon}, {Fynbo}, {Hambsch}, {Hamuy}, {Harutyunyan},
  {Ivarson}, {Kankare}, {Martin}, {Kotak}, {LaCluyze}, {Maguire}, {Mattila},
  {Maza}, {McCrum}, {Miluzio}, {Norgaard-Nielsen}, {Nysewander}, {Ochner},
  {Pan}, {Pumo}, {Reichart}, {Tan}, {Taubenberger}, {Tomasella}, {Turatto}, \&
  {Wright}}]{Pastorello13_09ip}
{Pastorello}, A., {Cappellaro}, E., {Inserra}, C., {et~al.} 2013, \apj, 767, 1

\bibitem[{{Pastorello} {et~al.}(2016){Pastorello}, {Wang}, {Ciabattari},
  {Bersier}, {Mazzali}, {Gao}, {Xu}, {Zhang}, {Tokuoka}, {Benetti},
  {Cappellaro}, {Elias-Rosa}, {Harutyunyan}, {Huang}, {Miluzio}, {Mo},
  {Ochner}, {Tartaglia}, {Terreran}, {Tomasella}, \&
  {Turatto}}]{2016MNRAS.456..853P}
{Pastorello}, A., {Wang}, X.-F., {Ciabattari}, F., {et~al.} 2016, \mnras, 456,
  853

\bibitem[{{Quataert} {et~al.}(2016){Quataert}, {Fern{\'a}ndez}, {Kasen},
  {Klion}, \& {Paxton}}]{2016MNRAS.458.1214Q}
{Quataert}, E., {Fern{\'a}ndez}, R., {Kasen}, D., {Klion}, H., \& {Paxton}, B.
  2016, \mnras, 458, 1214

\bibitem[{{Quataert} \& {Shiode}(2012)}]{2012MNRAS.423L..92Q}
{Quataert}, E., \& {Shiode}, J. 2012, \mnras, 423, L92

\bibitem[{Riemann(1860)}]{Riemann1860}
Riemann, B. 1860, Abhandlungen der K{\"o}niglichen Gesellschaft der
  Wissenschaften in G{\"o}ttingen, 8, 43

\bibitem[{Ro {et~al.}(2019)Ro, Coughlin, \& Quataert}]{cqr3}
Ro, S., Coughlin, E.~R., \& Quataert, E. 2019, The Astrophysical Journal, 878,
  150

\bibitem[{{Ro} \& {Matzner}(2013)}]{RoMatzner13_shocks}
{Ro}, S., \& {Matzner}, C.~D. 2013, \apj, 773, 79

\bibitem[{{Ro} \& {Matzner}(2017)}]{2017ApJ...841....9R}
---. 2017, \apj, 841, 9

\bibitem[{Ro(2017)}]{ro2017launching}
Ro, S.~S. 2017, PhD thesis, University of California, Berkeley

\bibitem[{{Sakurai}(1960)}]{sakurai60}
{Sakurai}, A. 1960, Comm. Pure Appl. Math, 13

\bibitem[{{Salbi} {et~al.}(2014){Salbi}, {Matzner}, {Ro}, \&
  {Levin}}]{Salbi14_Oblique}
{Salbi}, P., {Matzner}, C.~D., {Ro}, S., \& {Levin}, Y. 2014, \apj, 790, 71

\bibitem[{{Shiode} \& {Quataert}(2014)}]{2014ApJ...780...96S}
{Shiode}, J.~H., \& {Quataert}, E. 2014, \apj, 780, 96

\bibitem[{{Shivvers} {et~al.}(2016){Shivvers}, {Zheng}, {Mauerhan}, {Kleiser},
  {Van Dyk}, {Silverman}, {Graham}, {Kelly}, {Filippenko}, \&
  {Kumar}}]{2016MNRAS.461.3057S}
{Shivvers}, I., {Zheng}, W.~K., {Mauerhan}, J., {et~al.} 2016, \mnras, 461,
  3057

\bibitem[{{Shivvers} {et~al.}(2017){Shivvers}, {Zheng}, {Van Dyk}, {Mauerhan},
  {Filippenko}, {Smith}, {Foley}, {Mazzali}, {Kamble}, {Kilpatrick},
  {Margutti}, {Yuk}, {Graham}, {Kelly}, {Andrews}, {Matheson}, {Wood-Vasey},
  {Ponder}, {Brown}, {Chevalier}, {Milisavljevic}, {Drout}, {Parrent},
  {Soderberg}, {Ashall}, {Piascik}, \& {Prentice}}]{2017MNRAS.471.4381S}
{Shivvers}, I., {Zheng}, W., {Van Dyk}, S.~D., {et~al.} 2017, \mnras, 471, 4381

\bibitem[{{Smith}(2008)}]{2008Natur.455..201S}
{Smith}, N. 2008, \nat, 455, 201

\bibitem[{{Smith} \& {Arnett}(2014)}]{2014ApJ...785...82S}
{Smith}, N., \& {Arnett}, W.~D. 2014, \apj, 785, 82

\bibitem[{{Smith} {et~al.}(2011){Smith}, {Li}, {Filippenko}, \&
  {Chornock}}]{2011MNRAS.412.1522S}
{Smith}, N., {Li}, W., {Filippenko}, A.~V., \& {Chornock}, R. 2011, \mnras,
  412, 1522

\bibitem[{{Smith} {et~al.}(2014){Smith}, {Mauerhan}, \&
  {Prieto}}]{2014MNRAS.438.1191S}
{Smith}, N., {Mauerhan}, J.~C., \& {Prieto}, J.~L. 2014, \mnras, 438, 1191

\bibitem[{{Smith} {et~al.}(2012){Smith}, {Mauerhan}, {Silverman},
  {Ganeshalingam}, {Filippenko}, {Cenko}, {Clubb}, \&
  {Kandrashoff}}]{2012MNRAS.426.1905S}
{Smith}, N., {Mauerhan}, J.~C., {Silverman}, J.~M., {et~al.} 2012, \mnras, 426,
  1905

\bibitem[{{Smith} {et~al.}(2018){Smith}, {Rest}, {Andrews}, {Matheson},
  {Bianco}, {Prieto}, {James}, {Smith}, {Strampelli}, \&
  {Zenteno}}]{2018MNRAS.480.1457S}
{Smith}, N., {Rest}, A., {Andrews}, J.~E., {et~al.} 2018, \mnras, 480, 1457

\bibitem[{{Tan} {et~al.}(2001){Tan}, {Matzner}, \&
  {McKee}}]{2001ApJ...551..946T}
{Tan}, J.~C., {Matzner}, C.~D., \& {McKee}, C.~F. 2001, \apj, 551, 946

\bibitem[{Tyagi \& Sujith(2005)}]{tyagi2005propagation}
Tyagi, M., \& Sujith, R. 2005, Physica D: Nonlinear Phenomena, 211, 139

\bibitem[{Tyagi \& Sujith(2003)}]{tyagi2003nonlinear}
Tyagi, M., \& Sujith, R.~I. 2003, Journal of Fluid Mechanics, 492, 1

\bibitem[{{Ulmschneider}(1970)}]{1970SoPh...12..403U}
{Ulmschneider}, P. 1970, \solphys, 12, 403

\bibitem[{Whitham(1974)}]{whitham1974linear}
Whitham, G. 1974, Linear and Nonlinear Waves, A Wiley-Interscience publication
  (Wiley)

\bibitem[{Woosley(2017)}]{woosley2017pulsational}
Woosley, S. 2017, The Astrophysical Journal, 836, 244

\bibitem[{{Woosley} {et~al.}(2007){Woosley}, {Blinnikov}, \&
  {Heger}}]{2007Natur.450..390W}
{Woosley}, S.~E., {Blinnikov}, S., \& {Heger}, A. 2007, \nat, 450, 390

\bibitem[{{Woosley} \& {Heger}(2015)}]{2015ApJ...810...34W}
{Woosley}, S.~E., \& {Heger}, A. 2015, \apj, 810, 34

\bibitem[{{Wyman} {et~al.}(2004){Wyman}, {Chernoff}, \&
  {Wasserman}}]{WymanChernoffWasserman04}
{Wyman}, M.~C., {Chernoff}, D.~F., \& {Wasserman}, I. 2004, \mnras, 354, 1053

\end{thebibliography}

\end{document}